\documentclass[aps,prl,twocolumn,showpacs,amsmath,amssymb,floatfix,superscriptaddress,notitlepage]{revtex4-1}

\usepackage{color}
\usepackage{bbm}
\usepackage{graphicx}% Include figure files
\usepackage{dcolumn}% Align table columns on decimal point
\usepackage[utf8]{inputenc}
\usepackage{epstopdf}
\usepackage{amsthm}
\usepackage{amsmath}
\usepackage{empheq}
\usepackage{bbm}
\usepackage{braket}
\usepackage{amssymb}
\usepackage[usenames,dvipsnames]{xcolor}
\usepackage{cases}
\usepackage{latexsym}
\usepackage[colorlinks=true,citecolor=Cerulean,linkcolor=RubineRed,urlcolor=Cerulean]{hyperref}

\graphicspath{ {images/}{images/Figs-07-23-18/}{images/figure1/}{images/figure1/finales/}{images/figure2/}{images/figure3/}{images/figure3/finales/}{images/figure4/}{images/figure5/} }

\newcommand{\tr}[1]{\operatorname{\textnormal{Tr}}\left( {#1} \right)}  %Trace
\usepackage{lipsum}
\usepackage{graphicx}

\begin{document}

\title{Spontaneous symmetry breaking induced by quantum monitoring}
\author{Luis Pedro Garc\'ia-Pintos}
\thanks{Corresponding author}
\email{luis.garciapintos@umb.edu}
\affiliation{Department of Physics, University of Massachusetts, Boston, MA 02125, USA}
\author{Diego Tielas}
\email{tielas@fisica.unlp.edu.ar}
\affiliation{Department of Physics, University of Massachusetts, Boston, MA 02125, USA}
\affiliation{IFLP, CONICET - Department of Physics, University of La Plata, C.C. 67 (1900), La Plata, Argentina}
\affiliation{Faculty of Engineering, University of La Plata, La Plata, Argentina}
\author{Adolfo del Campo}
\email{adolfo.delcampo@umb.edu}
\affiliation{Donostia International Physics Center,  E-20018 San Sebasti\'an, Spain}
\affiliation{IKERBASQUE, Basque Foundation for Science, E-48013 Bilbao, Spain}
\affiliation{Department of Physics, University of Massachusetts, Boston, MA 02125, USA}
\affiliation{Theoretical Division, Los Alamos National Laboratory, MS-B213, Los Alamos, NM 87545, USA}
\date{\today}

\begin{abstract}

Spontaneous symmetry breaking (SSB) is responsible for structure formation in scenarios ranging from condensed matter to cosmology. SSB is broadly understood in terms of perturbations to the Hamiltonian governing the dynamics or to the state of the system. We study SSB due to quantum monitoring of a system via continuous quantum measurements. The acquisition of information during the measurement process induces a measurement back-action that seeds SSB. In this setting, by monitoring different observables,  an observer can tailor the topology of the vacuum manifold, the pattern of symmetry breaking and the nature of the resulting domains and topological defects.

\end{abstract}

\maketitle

Spontaneous symmetry breaking (SSB) occupies a central stage in modern physics~\cite{arodz2012patterns}. 
It governs physical mechanisms such as  BCS superconductivity, the existence of Nambu-Goldstone bosons~\cite{NambuPRL60,Goldstone61}, and the generation of mass via the Higgs mechanism~\cite{EnglertPRL64,HiggsPRL64,KibblePRL64}. It is also considered to play a key role in the early history of the universe~\cite{AdlerRevModPhys82,AlbrechtPRL82}, and may have contributed to structure formation in this context \cite{Kibble76a}.
Symmetry breaking arises in Nature when observable phenomena lack the symmetry of the underlying physical laws dictating them. A system can thus be found in configurations in conflict with the invariance manifested by its equations of motion. 

SSB is characteristic of phase transitions from a high-symmetry phase to a lower symmetry phase in which a new kind of macroscopic order emerges, that can be detected by an order parameter~\cite{Sachdev}. In a specific system, SSB is analyzed through the  symmetries of its free energy landscape or field equations~\cite{Kibble76a}.
An intuitive understanding of the spontaneous rupture of symmetry is acquired by picturing a spin chain initially prepared in a paramagnetic phase and driven by a Hamiltonian $H \propto -\sum_j \sigma^z_j \sigma^z_{j+1}$, which favors ferromagnetic order. 
The ensuing dynamics preserves the symmetries of the Hamiltonian, and in the absence of external perturbations nothing in the evolution biases the choice between the degenerate ground states $\ket{ \uparrow \uparrow \dots \uparrow \uparrow}$ and $\ket{ \downarrow \downarrow \dots \downarrow \downarrow}$.
However, in the thermodynamic limit any infinitesimal external magnetic field is enough to break the tie, and single out one of these states as the observed ground state. In this case the symmetry is explicitly broken. 
In physical systems of finite size, the breaking of the symmetry can be 
generally understood as a consequence of random fluctuations, either in the Hamiltonian governing the evolution, or in the state of the system at a given time. 
In the example of the spin chain driven from the symmetric paramagnetic phase,
a local perturbation of the magnetization is enough to implant a seed that leads to symmetry breaking along the whole system. 
Whenever present, spatial fluctuations of the magnetization can thus favor the  local growth of domains where spins align in a certain direction.
This example serves to illustrate the  general physical mechanism of domain formation. 

A characteristic feature of the canonical description of  SSB is its focus on static features: it is understood via the properties of the state of the system in thermal equilibrium. This approach is frequently pursued using elements of homotopy theory to characterize the topology of the vacuum manifold, spanned by the degenerate ground states in the broken-symmetry phase \cite{Kibble76a,MerminRevModPhys79}. 

In this 
article we 
study an alternative mechanism for symmetry breaking, induced by the continuous monitoring of a quantum system. 
%In this context, SSB requires no ad-hoc perturbations to the Hamiltonian or to the state of the system, this approach is dynamical and purely quantum mechanical in nature. 
During the monitoring by continuous quantum measurements, symmetry breaking is induced by the quantum measurement back-action~\cite{DevoretSCIENCE2013} associated with the  acquisition of information by the observer. The latter can thus alter the topology of the vacuum manifold and the pattern of symmetry breaking.  By selecting different kinds of measurements, an observer can thus control the nature of the resulting domains and topological defects, and even achieve a complete suppression of the later.

\emph{Monitoring-induced symmetry breaking ---}
For the sake of illustration, we consider $N$ $1/2$-spins  prepared in the ground state of a paramagnetic Hamiltonian $H_0 =- \Lambda \sum_{j=1}^{N} \sigma_j^x $, where $\Lambda$ represents a global energy scale.
The initial state of the chain is then 
\begin{align}
\ket{\Psi(0)} = \bigotimes_{j=1}^{N} \ket{\rightarrow}_j,
\end{align}
where $\ket{\rightarrow}_j$ denotes the eigenstate of $\sigma_j^x$ with eigenvalue $1$. 
Following  a sudden quench at $t=0$, the system evolves for $t > 0$ according to the ferromagnetic Hamiltonian
\begin{equation}
\label{eq:hamiltonian}
H = - \Lambda  \sum_{j=1}^{N} \sigma_j^z\sigma_{j+1}^z ,
\end{equation}
taking periodic boundary conditions $\sigma_{N+1}^z \equiv \sigma_1^z$ for concreteness.

The initial state of the chain shares a symmetry of the Hamiltonian, which is preserved in the case of unitary evolution. 
This can be easily seen, for instance, by noting that 
the Hamiltonian commutes with the magnetization $M = \sum_j \sigma_j^z$ of the chain. % i.e. $[H,M] = 0$.
 In the adiabatic limit, since the initial state satisfies $\bra{ \Psi(0)} M \ket{\Psi(0)} = 0$,
the subsequent evolved state necessarily has equal weights on the states $\ket{ \uparrow \uparrow \dots \uparrow \uparrow}$ and $\ket{ \downarrow \downarrow \dots \downarrow \downarrow}$, and is therefore incapable of selecting between them.
However, experience shows that in practice a particular direction for the chain's magnetization is spontaneously chosen by the system, even if Hamiltonian and initial state are symmetric with respect to the $z$ direction. In a sudden quench scenario, excitations in the system are constrained by the conservation law $\bra{ \Psi(t)} M \ket{\Psi(t)} = 0$.
The symmetry is broken locally with the formation of domains in which spins are homogeneously polarized along a given direction (up or down) and the formation of topological defects at the interface between adjacent domains. In the example at hand, the latter are the so-called $\mathbb{Z}_2$-kinks, as the broken $\mathbb{Z}_2$ symmetry is restored at their core. In the canonical approach to symmetry breaking, domain formation is explained by assuming a small perturbation to the Hamiltonian or to the state of the system, locally breaking the symmetry `by hand'.

We consider an alternative SSB scenario that results from monitoring a quantum system  during time evolution.
Such quantum monitoring can be modeled by continuous quantum measurements~\cite{JacobsIntro2006,Bookwiseman2009,Bookjacobs2014}, which can be thought of as a sequence of infinitesimally weak measurements.
More specifically, they arise as the consequence of a weak coupling between the system being measured and an apparatus that gets entangled with the state of the system. Upon observing a particular outcome in the measurement apparatus the joint state is collapsed.
In contrast to strong projective measurements, which can drastically perturb the state of the system, a weak measurement provides only partial information of the state. 
In doing so, this process induces a mild back-action on the state of the system at any given time.
The collective information obtained from the continuous measurement record over a period of time can provide full information of the system though~\cite{SiddiqiNature2014}, and can thus serve as a way to perform full state tomography~\cite{SmithPRL2006,MurchNature2013}, parameter estimation~\cite{MabuchiPRL2002,MolmerPRL2014}, quantum error correction~\cite{AhnPRA2002}, and quantum control~\cite{DohertyPRA1999}.
Note that the action of the observer is in practice tantamount to the coupling to a monitoring environment, whenever the latter weakly interacts with the system of interest in such a way that information of a physical quantity is probed and registered~\cite{Zurek1992}. 
The relation between quantum dynamics, decoherence, and SSB have been studied in the past~\cite{vanWezelPRL2005,Dziarmaga12}.
Particular aspects of the connection between quantum measurement and symmetry breaking have been considered, mostly as an effective description to explain the interference fringes in  Bose-Einstein condensates  and superfluids \cite{Leggett1991,CiracPRA1996,JavanienPRL1996,CastinPRA1997,  PARKINS1998PhysicsReports,MiyaderaPRL2002,Paraoanu2008JLTP}.
In what follows we focus on the agency of the observer to control symmetry breaking via the selection of the continuous measurement, i.e., the observable that is monitored. In particular, we shall discuss symmetry breaking induced by the monitoring of  local, coarse-grained and global observables.

\begin{figure} 
  \vspace{-42pt} 
  \centering
   \includegraphics[angle=90,origin=c,width=0.45\textwidth]{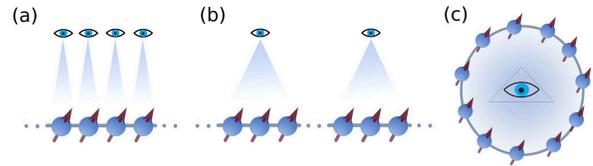} 
        \vspace{-82pt} 
\caption{
\label{fig:figure2} 
{\bf 
From single spin to coarse-grained monitoring.
}
The nature of the monitoring process determines the pattern of measurement-induced symmetry breaking, 
deciding  the size of the domains, the distribution of the local magnetization, and the type of topological defects formed in the system.
(a) Independent observers measuring each single spin induce random, independent measurement outcomes for each spin, which end up either `up' or `down'.
(b) Independent observers measuring the magnetization of clusters of spins provoke the formation of local and independent domains, which can take different values of the magnetization.  
(c) In the limiting case of one single observer measuring the whole spin chain, the induced global magnetization due to quantum monitoring is homogeneous.
}
\end{figure}

\begin{figure*}%[ht]
  \centering
        \includegraphics[trim=00 00 00 00,width=0.27\textwidth]{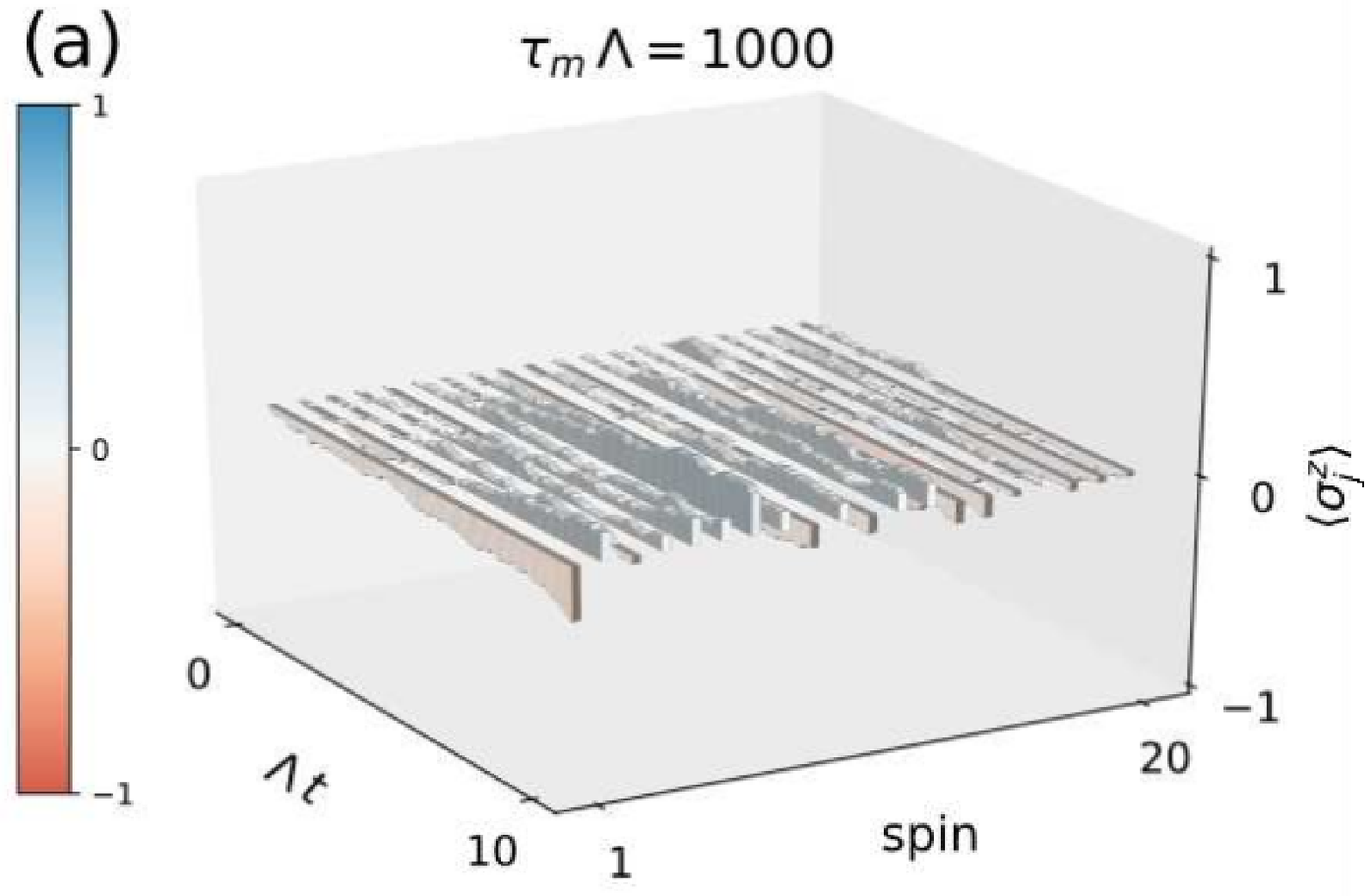}
%  \includegraphics[trim=00 00 00 00,width=0.40\textwidth]{QCM_3D_N20-tm100-Nstep256-tq10.eps}
%  \\
         \includegraphics[trim=00 00 00 00,width=0.27\textwidth]{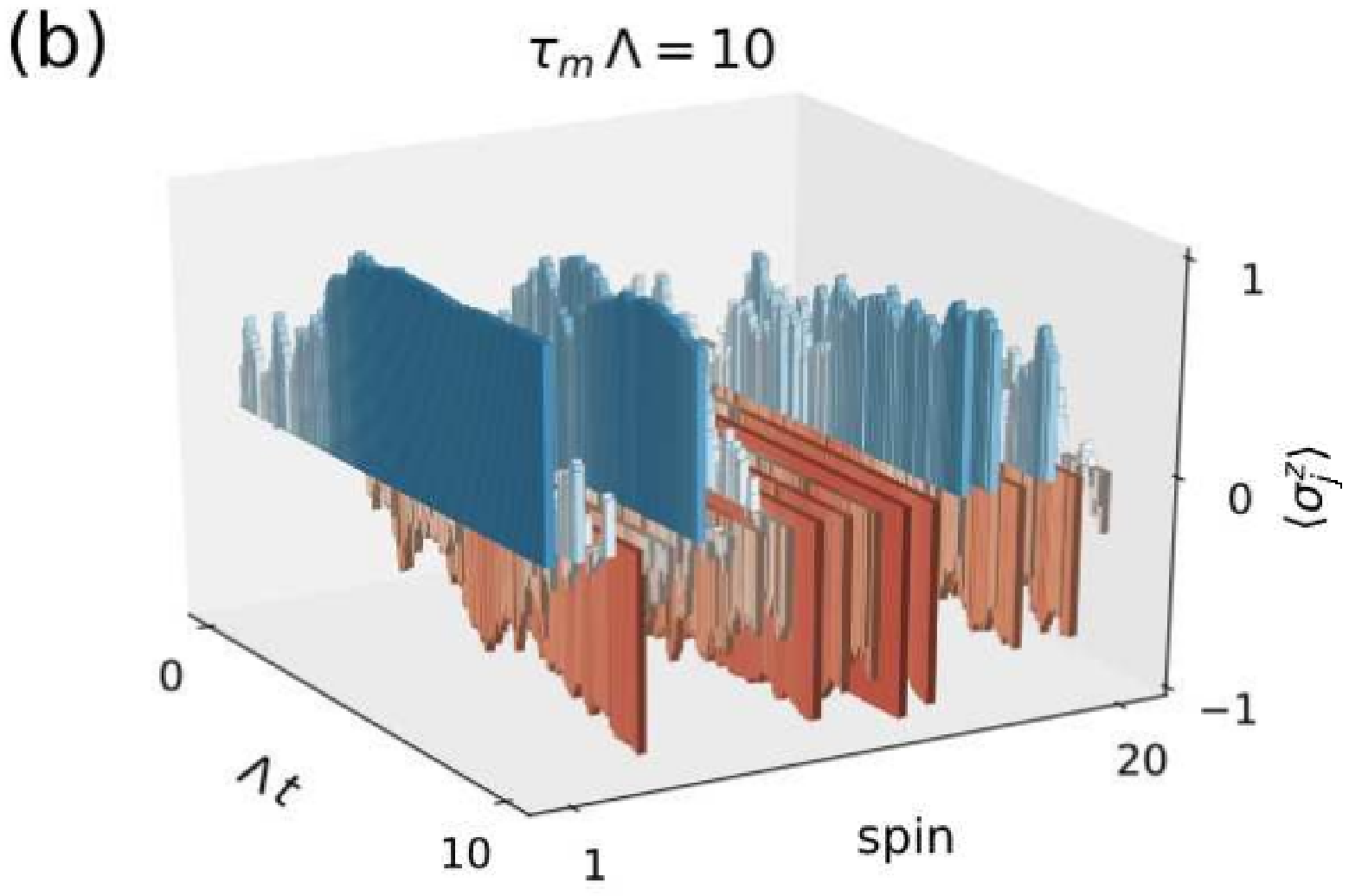}  
       \includegraphics[trim=00 00 00 00,width=0.27\textwidth]{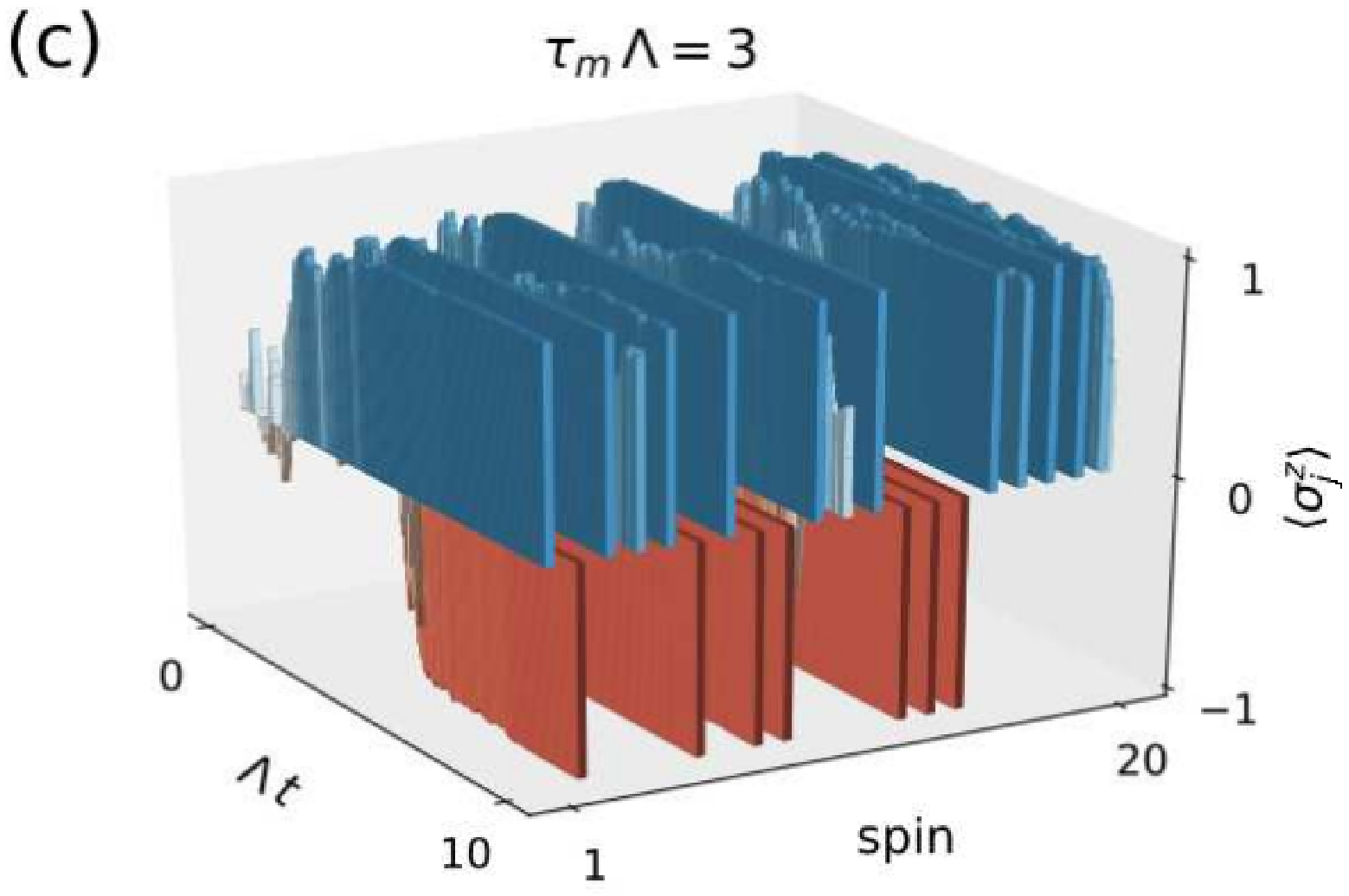}  
\caption{
\label{fig:figure1} 
{\bf 
Dynamics of symmetry breaking  induced by monitoring of each spin for different measurement strengths.}
(a) A weak monitoring of the individual spins slightly affects their spin components, barely perturbing the otherwise symmetric evolution. 
(b) As the measurement strength is increased symmetry is broken by the measurement process, leading to spins collapsing to the stable `up' or `down' configurations.
(c) A strong monitoring process rapidly leads to the formation of localized topological defects in individual realizations of the stochastic evolution. 
}
\end{figure*}

The dynamics of the system undergoing continuous monitoring of an arbitrary set of observables $\{A_\alpha\}$ is well described by a stochastic master equation dictating the change in the state,
\begin{align}
\label{eq:changerho}
d \rho_t &= L\left[\rho_t\right] dt + \sum_\alpha I_\alpha \left[\rho_t\right] dW^\alpha_t ,
\end{align}
when expressed in It\^o form \cite{Bookwiseman2009,Bookjacobs2014}. Here,  $L[\rho_t]$ takes the standard Lindblad form for the set of measured operators, which includes the evolution due to the Hamiltonian and dephasing due to the monitoring process:  
\begin{align}
\label{eq:Lindblad}
L[\rho_t] &=-i \left[H,\rho_t \right] -   \sum_\alpha \frac{1}{8\tau_m^\alpha} \left[A_\alpha,\left[A_\alpha,\rho_t \right]\right]. 
\end{align}
In turn, the `innovation terms' 
\begin{align}
\label{eq:innovation}
I_\alpha \left[\rho_t\right] &= \sqrt{\frac{1}{4\tau_m^\alpha}} \left(\left\{ A_\alpha ,\rho_t \right\} - 2\tr{A_\alpha \rho_t} \rho_t \right) 
\end{align}
account for the change in the state of the system due to the acquisition of information during the measurement process. 
These innovation terms %, nonlinear in $\rho_t$, 
encompass the effect of the back-action on the state of the system due to the quantum measurement. 
%It can be shown that 
Here $dW^\alpha_t$ denote independent Gaussian random variables of mean $0$ and width $dt$, while  $\tau_m^\alpha$ is the `characteristic measurement time'
 with which observable $A_\alpha$ is monitored, i.e., it provides the timescales over which information of the expectation value of the observable is acquired. % in the measurement process. 
The output of such measurements over an interval $dt$, given by $dr_\alpha(t)~=~\langle A_\alpha \rangle(t)dt~+~\sqrt{\tau_m^\alpha} dW_t^\alpha $, provides information of the expectation value of the observables, hidden by additive white noise~\cite{JacobsIntro2006,
Bookwiseman2009,Bookjacobs2014}.
%For simplicity we have taken units such that $\hbar = 1$.
The time evolution is unravelled by modeling continuous measurements with a sequence of infinitesimally weak measurements, which can be described by Kraus operators acting on the state at every time step $dt$ of the evolution, see for example~\cite{JacobsIntro2006}.  In the limit of $dt$ smaller than any other relevant timescales, such approach gives the same dynamics as equation~\eqref{eq:changerho}. All simulations are performed in QuTiP~\cite{qutip2012,qutip2013}.

We shall focus on the local, coarse-grained and global magnetization as choice of monitored observables.
To start with, assume that independent observers continuously monitor the single-spin components $\{ \sigma_j^z \}$ of  individual spins along the chain, as illustrated in Fig.~\ref{fig:figure2}(a). 
The dynamics is %then
 dictated by equation~\eqref{eq:changerho}, with $\{ \sigma_j^z \}$ ($j=1,\dots,N$) as the set of measured observables $\{A_\alpha\}$.

This scenario  makes apparent the connection between the dynamics under continuous measurements and that of an open system in contact with an environment.
Observers without access to the measurement output, who need to average over the unobserved measurement outcomes, obtain an averaged description of the state of the system $\rho_t^m$. The latter  evolves according to $d \rho_t^{m}~=~L\left[\rho_t^m\right] dt$, and its evolution is thus  identical to that of the system coupled to an environment through the spin components $\{ \sigma_j^z \}$. 
Importantly, such density matrix does not show signs of symmetry breaking, given that the evolution commutes with the
magnetization. %parity operator.
That is, without registering the measurement outcomes and in the absence of any further perturbations to Hamiltonian or state, symmetry is fully preserved. In particular, the spin components $\langle \sigma_j^z \rangle(t) = \tr{\rho_t^m \sigma_j^z}$ remain constant.

By contrast, the measurement process does break the symmetry, forcing individual spins to collapse to one of the eigenstates of $\sigma_j^z$.
Indeed, when conditioning the state to the observed outcomes the measurement back-action breaks the symmetry in individual realizations.
To prove this, let us focus on the evolution of the component of spin $j$ in the ferromagnetic part of the quench, with $t \ge 0$. 
%since the Hamiltonian commutes with $\sigma_j^z$ in this regime, 
The evolution of the expectation value of the spin components % $\sigma_j^z$
  is dictated by
\begin{align}
\label{eq:changeexpectation}
 d \langle \sigma_j^z \rangle(t) &= -\tr{ I \left[\rho_t \right]  \sigma_j^z} dW^j_t \nonumber \\
&= -\sqrt{\frac{1}{\tau_m}} \Delta^2_{\sigma_j^z}(t) dW_t^j,
\end{align}
where  the trace $\langle \cdot \rangle(t) \equiv \tr{\rho_t \cdot }$  is taken with respect to the state $\rho_t$, and $\Delta_{\sigma_j^z}(t) = \sqrt{\left\langle \left( \sigma_j^z \right)^2 \right\rangle(t) - \langle \sigma_j^z \rangle^2(t)}$ is the corresponding standard deviation.
This means that, due to quantum monitoring, the spin component evolves whenever its quantum uncertainty is non-zero. Such uncertainty is zero if and only if the spin is in one of the eigenstates $\ket{\uparrow}_j$ or $\ket{\downarrow}_j$ of $\sigma_j^z$. 
Therefore, only states with definite values of the spin component are stable under monitoring. 
SSB is thus a consequence of  the measurement back-action encoded in the `innovation terms' $I_j[\rho_t]$ in equation~\eqref{eq:innovation}. 

Effectively, monitoring the spin components breaks the symmetry in the chain, in the basis selected by the measurement process, as illustrated in Fig.~\ref{fig:figure1}.
The characteristic measurement time $\tau_m$ dictates the rate at which symmetry breaking occurs and individual spins collapse `up' or `down'.
The stochastic dynamics naturally leads to the formation of localized topological defects~\cite{MerminRevModPhys79},  $\mathbb{Z}_2$-kinks, from the sole effect of the quantum measurement back-action due to the monitoring of the spin chain.

\begin{figure}%[ht]
  \centering
        \includegraphics[trim=00 00 00 00,width=0.238\textwidth]{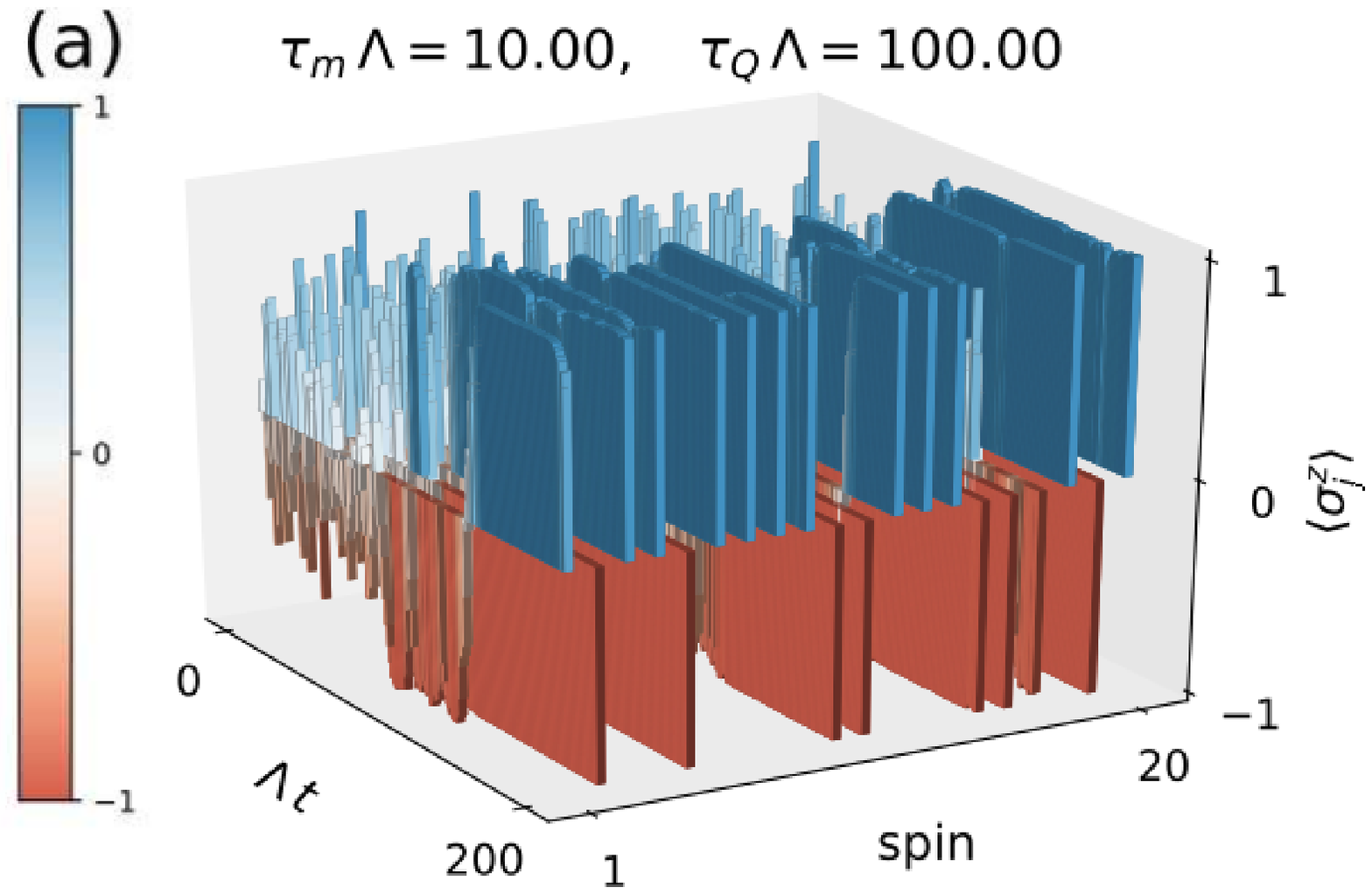}
%  \includegraphics[trim=00 00 00 00,width=0.40\textwidth]{QCM_3D_N20-tm100-Nstep256-tq10.eps}
%  \\
%         \includegraphics[trim=00 00 00 00,width=0.27\textwidth]{figQCMquenched_3D_N20-Nstep256-Nclusters20-tm50-tq25_V2.eps}  
       \includegraphics[trim=00 00 00 00,width=0.238\textwidth]{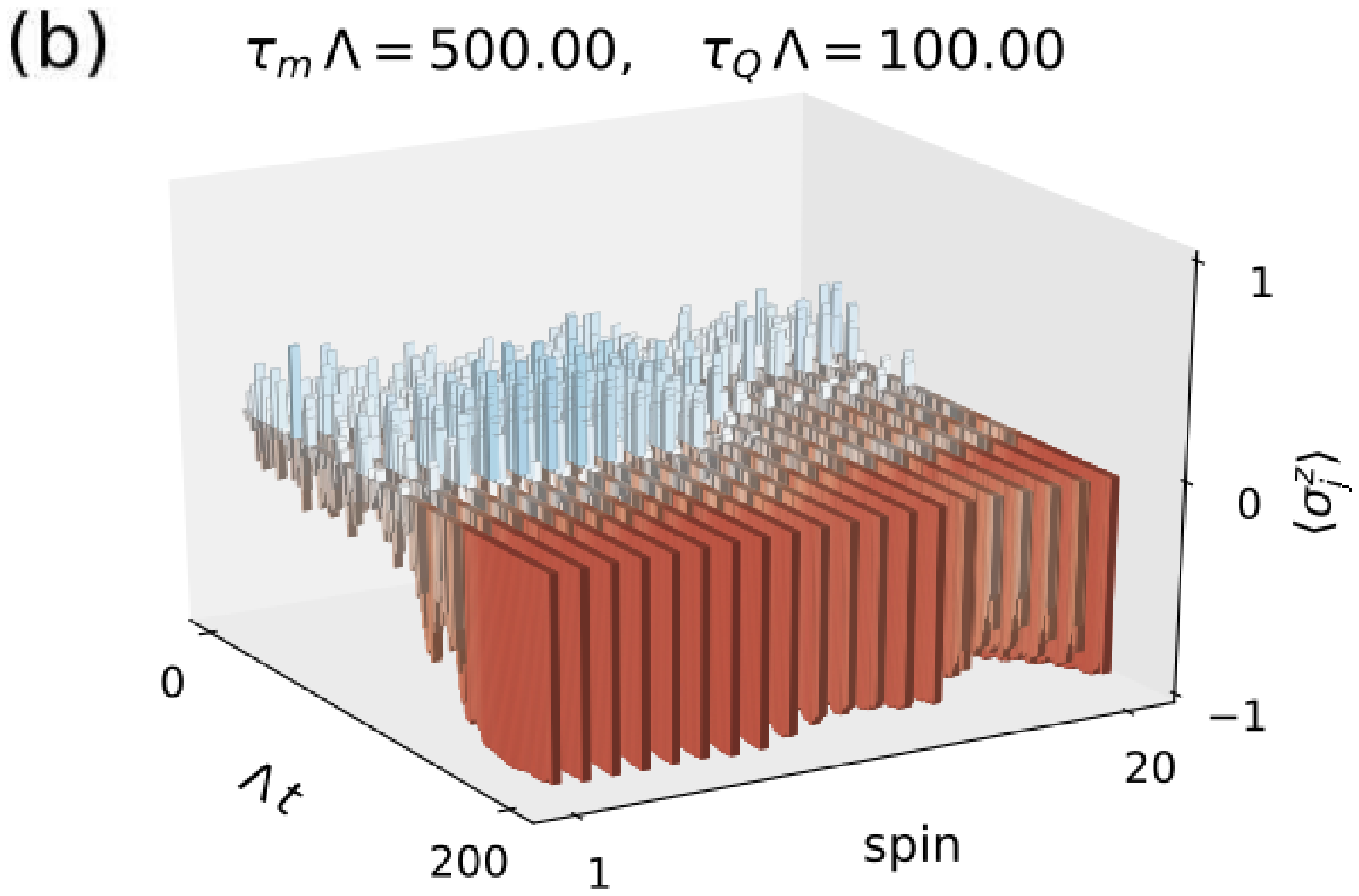}  
\caption{
\label{fig:figure4} 
{\bf Monitoring-induced symmetry breaking for a slow quench.}
A system undergoing a slow quench ($1/\tau_Q \ll \Lambda$) tends to remain close to the ground state, but in the absence of monitoring and other sources of symmetry breaking the system cannot select between degenerate ferromagnetic ground states $\ket{ \uparrow \uparrow \dots \uparrow \uparrow}$ and $\ket{ \downarrow \downarrow \dots \downarrow \downarrow}$.
(a) A strong monitoring of the individual spins ($1/\tau_m \sim 1/\tau_Q \ll \Lambda$) generates a symmetry broken state where spins are uncorrelated.
%slightly affects their spin components, barely perturbing the otherwise symmetric evolution. 
%(b) As measurement strength decreases, symmetry is broken without strongly affecting the tendency to remain close to the ground state. This favors formation of domains of spins collapsing to stable `up' or `down' configurations.
(b) A weak monitoring ($1/\tau_m \ll \{\Lambda,1/\tau_Q\}$) leads to an almost uniform domain, and a final state that is close to one of the possible symmetry-broken ground states of the ferromagnetic Hamiltonian. 
}
\end{figure}

The previous analysis applies to a sudden quench in the dynamics. We now consider the scenario of a slow transition from a paramagnetic to a ferromagnetic regime, with a time-dependent Hamiltonian
\begin{equation}
\label{eq:hamiltonianQuench}
H(t) = - \Lambda \left( 1- \frac{t}{\tau_Q} \right) \sum_{j=1}^{N} \sigma_j^x  - \Lambda \frac{t}{\tau_Q}  \sum_{j=1}^{N} \sigma_j^z\sigma_{j+1}^z ,
\end{equation}
where $\tau_Q$ is a timescale that dictates completion of the quench. % to the ferromagnetic Hamiltonian.
 For times $t~\ge~\tau_Q$ we assume $H(t) = H(\tau_Q)$. 
In contrast to sudden-quenches, the finite-time crossing of the critical point reduces excitations,  favoring dynamics constrained to the lowest energy subspace~\cite{Born1928}. However, without access to measurement outcomes nothing in the dynamics breaks the symmetry between degenerate grounds states of the ferromagnetic Hamiltonian. 
The monitoring of the system feeds in a seed of asymmetry, as Fig.~\ref{fig:figure4} illustrates. In this case there is a competition between monitoring back-action, which singles out individual spins, and the natural tendency of the system to remain excitation-less due to adiabatic dynamics. Domains with different definite values of magnetization form, with a size that depends on measurement strength and quench time (see Supplemental Material).

%\cite{calabrese2012Jstat}
%\begin{figure} 
%  \centering 
%     \includegraphics[trim=00 00 00 00,width=0.238\textwidth]{QCM_3Dcoarsegrained_N20-tm2-Nstep2560-tq10_V2.eps}
%    % \\
%     %\vspace{7pt}
%    \includegraphics[trim=00 00 00 00,width=0.238\textwidth]{QCM_3Dglobal_N20-tm2-Nstep2560-tq10_V2.eps}
%\caption{
%\label{fig:figure4} 
%\color{blue}
%A system undergoing a slow quench tends to remain close to the ground state. In the absence of monitoring and other sources of symmetry breaking the system cannot select between degenerate ferromagnetic ground states $\ket{ \uparrow \uparrow \dots \uparrow \uparrow}$ and $\ket{ \downarrow \downarrow \dots \downarrow \downarrow}$.
%XXXXXXXXXXXX
%(a) The continuous monitoring of the individual spin compon
%coarse-grained local magnetization breaks the symmetry, resulting in the formation of independent domains.
%For independent observers monitoring the local magnetization over the spin chain, the typical size of the domains is determined by the number of spins $K$ over which the independent measurement operators $m_\alpha$ act on. Moreover, the stable final states within each domain need not correspond to all spins pointing `up' or `down', as $m_\alpha$ can take $(K+1)$ different values.
%(b) In the extremal case of one observer monitoring the global magnetization, a single domain is formed.
%The nature of the monitoring process thus governs structure formation in the final state.
%}
%\end{figure}

The measured observable is also crucial in determining the nature of the symmetry breaking, and in particular, in governing structure formation in  the end state.
To illustrate this we analyze the case in which a coarse-grained local magnetization is probed on the chain.
Let us then consider that, 
instead of monitoring each individual spin component as in Fig. \ref{fig:figure2}(a), 
the local magnetization over clusters of $K$ consecutive spins is continuously measured, as on Fig.~\ref{fig:figure2}(b). 
We denote local magnetization observables by
\begin{align}
\label{eq:coarsegrainedobservable}
m_\alpha = \sum_{j \in \mathcal{I}_\alpha} \sigma_j^z, \qquad \alpha = \{1,\dots, N/K\},
\end{align}
% is performed, where the integer 
% $\alpha = \{1,\dots, N/N_c\}$ denotes , so 
where $\mathcal{I}_\alpha = [ K(\alpha-1) + 1 , K\alpha]$.

\begin{figure} 
  \centering 
     \includegraphics[trim=00 00 00 00,width=0.238\textwidth]{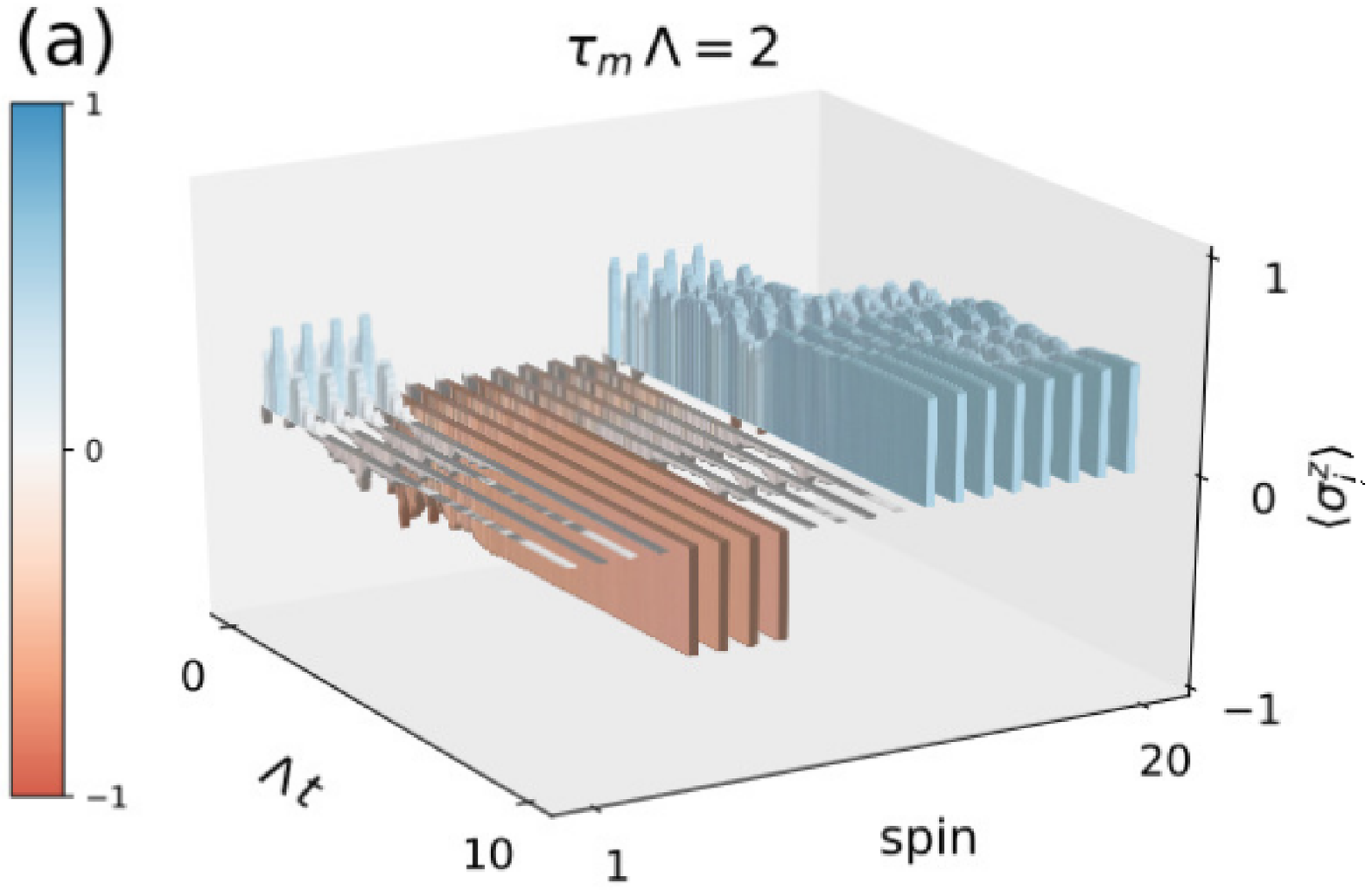}
    % \\
     %\vspace{7pt}
    \includegraphics[trim=00 00 00 00,width=0.238\textwidth]{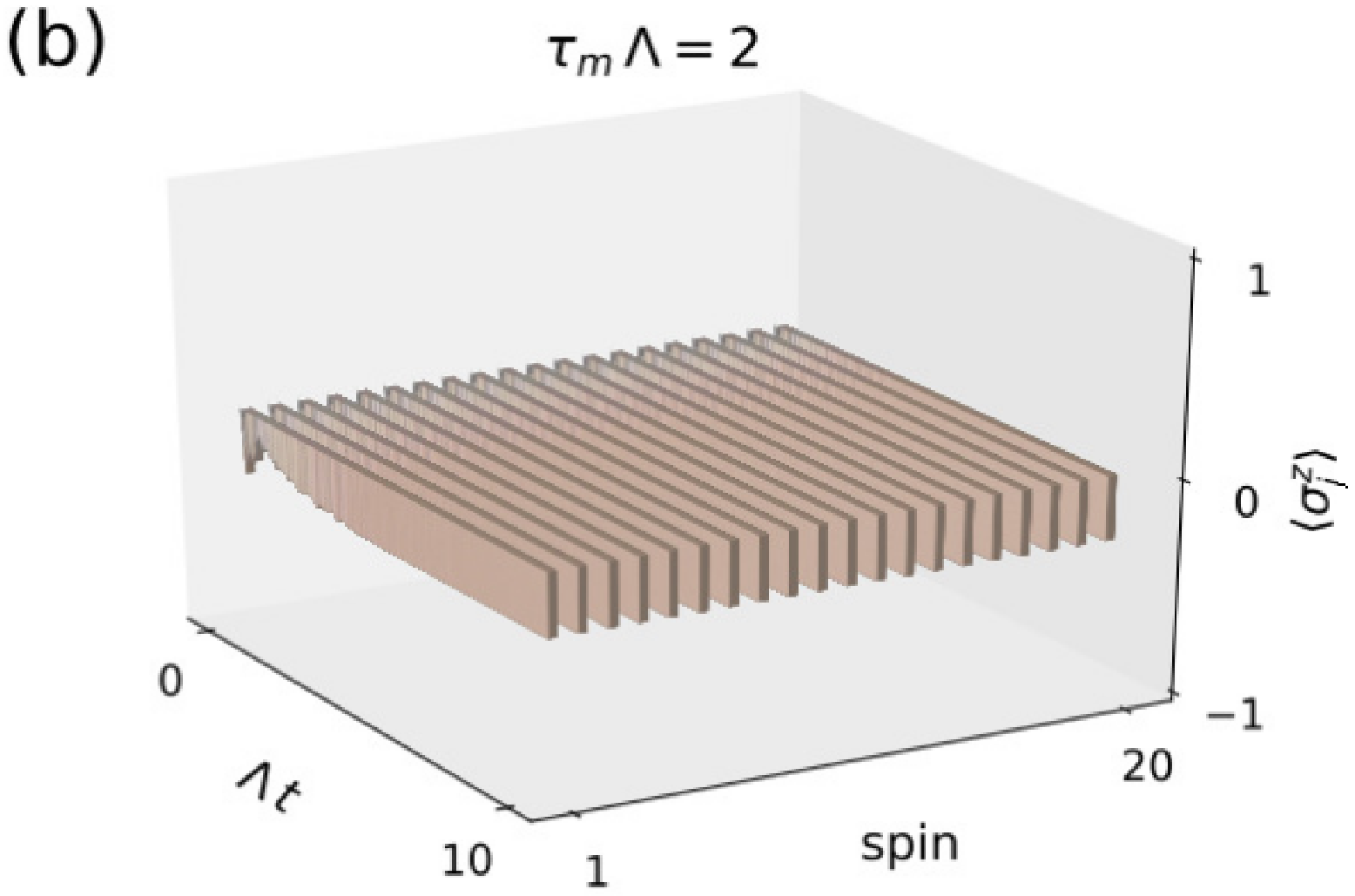}
\caption{
\label{fig:figure3} 
{\bf Evolution of $\langle \sigma_j^z \rangle$ for coarse-grained measurements.}
(a) The continuous monitoring of the coarse-grained local magnetization breaks the symmetry, resulting in the formation of independent domains.
For independent observers monitoring the local magnetization over the spin chain, the typical size of the domains is determined by the number of spins $K$ over which the independent measurement operators $m_\alpha$ act on. Moreover, the stable final states within each domain need not correspond to all spins pointing `up' or `down', as $m_\alpha$ can take $(K+1)$ different values.
(b) In the extremal case of one observer monitoring the global magnetization, a single domain is formed.
The nature of the monitoring process thus governs structure formation in the final state.
}
\end{figure}

Once again, symmetry is broken by the monitoring process, given that $d \langle m_\alpha \rangle(t) = -\sqrt{\frac{1}{\tau_m}} \Delta^2_{ m_\alpha }(t) dW_t^\alpha$, where $\Delta_{ m_\alpha }(t)$ denotes the standard deviation of the monitored coarse-grained magnetization.
In this case, the stable states to which the measurement process leads to, eigenvectors of the set of measured observables $\{ m_\alpha \}$, are starkly different, given that each spin cannot be singled out by the measurement process. 
Such eigenstates have definite values of the magnetization, $\lambda_m = \{-K, -K + 1, \dots, K - 1 ,  K\}$, on each of the coarse-grained regions. 
%As a result, t
This causes a symmetry breaking with non-uniform magnetization along the chain, but homogeneous behavior within individual spin clusters, as illustrated in Fig.~\ref{fig:figure3}(a). 
The resulting domains involve coherent quantum superpositions and have no classical counterpart. %, XXXexcept for the extreme case in which all $K$ spins point along the same directionXXX. 
Further, a topological defect  formed at the interface between such quantum domains is no longer restricted to  the type of $\mathbb{Z}_2$ kinks, but can result from a discontinuity in the local magnetization between any two of its $(K+1)$ possible values. Monitoring the coarse-grained magnetization broadens the class of topological defects to $\mathbb{Z}_{K+1}$ kinks, with the value of $K$ being controlled by the observers.

An interesting limiting case concerns the choice $K~=~N$,  that corresponds to  the monitoring of a single observable, the global magnetization $M = \sum_j \sigma_j^z$ of the spin chain, illustrated in Fig.\ref{fig:figure2}(c). 
In this scenario, the final state in the broken phase has very different properties from the case in which %various
 spin clusters are monitored. As shown in Fig.~\ref{fig:figure3}, it  results in a homogeneous magnetization along the chain, facilitating the growth of a single quantum domain, possibly including coherent quantum superpositions of spins with different components in the $z$-axis. In this case,  a complete suppression of topological defects is achieved.
By comparing the different choices of $\{A_\alpha\}$, we further conclude that the structure of states in the broken symmetry phase provides information of the nature of the monitoring process.

\emph{Discussion ---} 
We have studied the breaking of symmetry induced by the measurement back-action resulting from continuous measurements performed on the system of interest. In this context we have emphasized the agency of the observer to control symmetry breaking. In particular, we have analyzed the dynamics of a monitored many-body spin chain and identified strikingly different physics depending on the nature of the monitored observables. 
The individual monitoring of spins is consistent with a classical description and results in spins with definite `up' or `down' states, randomly assigned. By contrast, the use of local coarse-grained measurements of the local magnetization of clusters of spins leads to quantum domains characterized by coherent quantum superpositions, broadening the class of topological defects that the system can exhibit.  
Further, when a global magnetization is monitored, symmetry can be broken while the formation of topological defects is fully suppressed.
Thus, an observer can control the patterns of symmetry breaking by a choice of the measurement observables. This choice determines the nature of the final state in the broken symmetry phase, including the size and kind of domains, and the statistics of the magnetization. In such scenario, the classification of the resulting topological defects is no longer described by the symmetries of the system Hamiltonian, e.g. using homotopy theory \cite{Kibble76a,MerminRevModPhys79}.
The nature of the domains produced in this setting provides information of the monitoring process. In a broader context, the measurement back-action is expected to  govern  pattern formation.
Notably, the engineering of the patterns of symmetry breaking by continuous monitoring is amenable to experimental test in superconducting qubit platforms with current technology \cite{Houck12,Georgescu14}. 
It also opens up the possibility of using quantum control methods~\cite{WisemanJournalPhysB2002,
RoaPRA2006,
NeeleyScience2009,
JacobsNJphys2010,
WisemanNature2011,
DicarloNature2013,
MotzoiPRA2015} to tailor the end state in the  broken-symmetry phase.

\emph{Acknowledgements --- }
We thank I\~nigo L. Egusquiza, Fernando J. G\'omez-Ruiz, and Maxim Olshanii for comments on the manuscript. 
This work was partly funded by the John Templeton Foundation, UMass Boston (project P20150000029279), DOE grant DE-SC0019515, the National Research Council of Argentina (CONICET, PIP 12-2008-01-00282), and the Fulbright Visiting Scholar Program.

\bibliography{referencescontsymmbreak}

\section{Supplemental Material --- Spontaneous symmetry breaking induced by quantum monitoring}

\subsection{Monitoring-induced symmetry breaking under a slow quench}

We consider the dynamics of a continuously-monitored quantum phase transition  from a paramagnetic to a ferromagnetic regime in a timescale $\tau_Q$, as opossed to a sudden quench. The  time-dependent Hamiltonian reads
\begin{equation}
\label{eq:hamiltonianQuench}
H(t) = - \Lambda \left( 1- \frac{t}{\tau_Q} \right) \sum_{j=1}^{N} \sigma_j^x  - \Lambda \frac{t}{\tau_Q}  \sum_{j=1}^{N} \sigma_j^z\sigma_{j+1}^z.
\end{equation}
For times $t~\ge~\tau_Q$ we take $H(t) = H(\tau_Q)$. 
By contrast to sudden-quenches, the finite-time crossing of the critical point reduces excitations and instead favors dynamics constrained to the lowest energy subspace~\cite{Born1928}. 

\begin{figure*}  %[ht]
   \centering
        \includegraphics[trim=00 00 00 00,width=0.27\textwidth]{figQCMquenched_3D_N20-Nstep256-Nclusters20-tm10-tq100.eps}
%  \includegraphics[trim=00 00 00 00,width=0.40\textwidth]{QCM_3D_N20-tm100-Nstep256-tq10.eps}
%  \\
         \includegraphics[trim=00 00 00 00,width=0.27\textwidth]{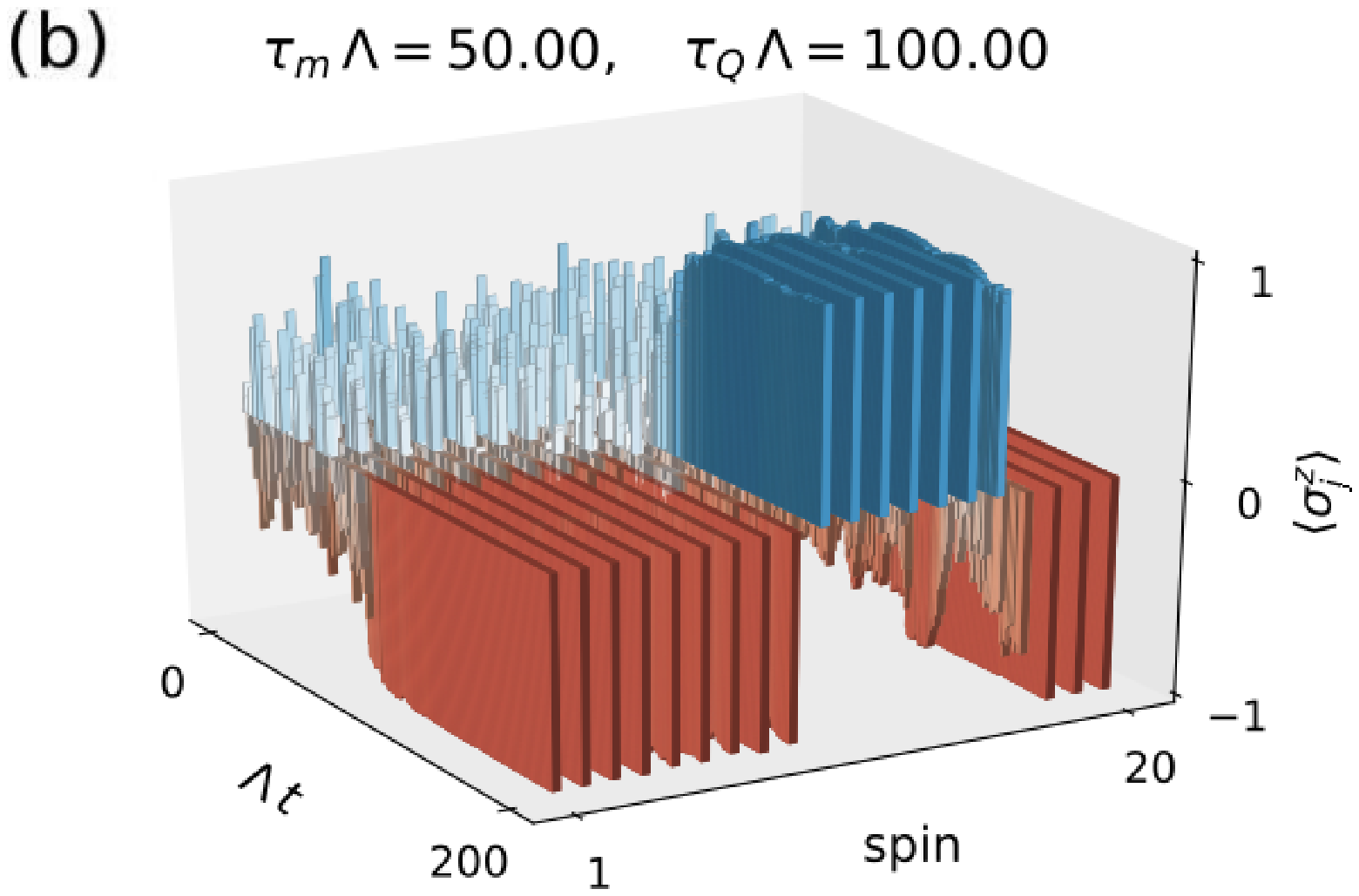}  
       \includegraphics[trim=00 00 00 00,width=0.27\textwidth]{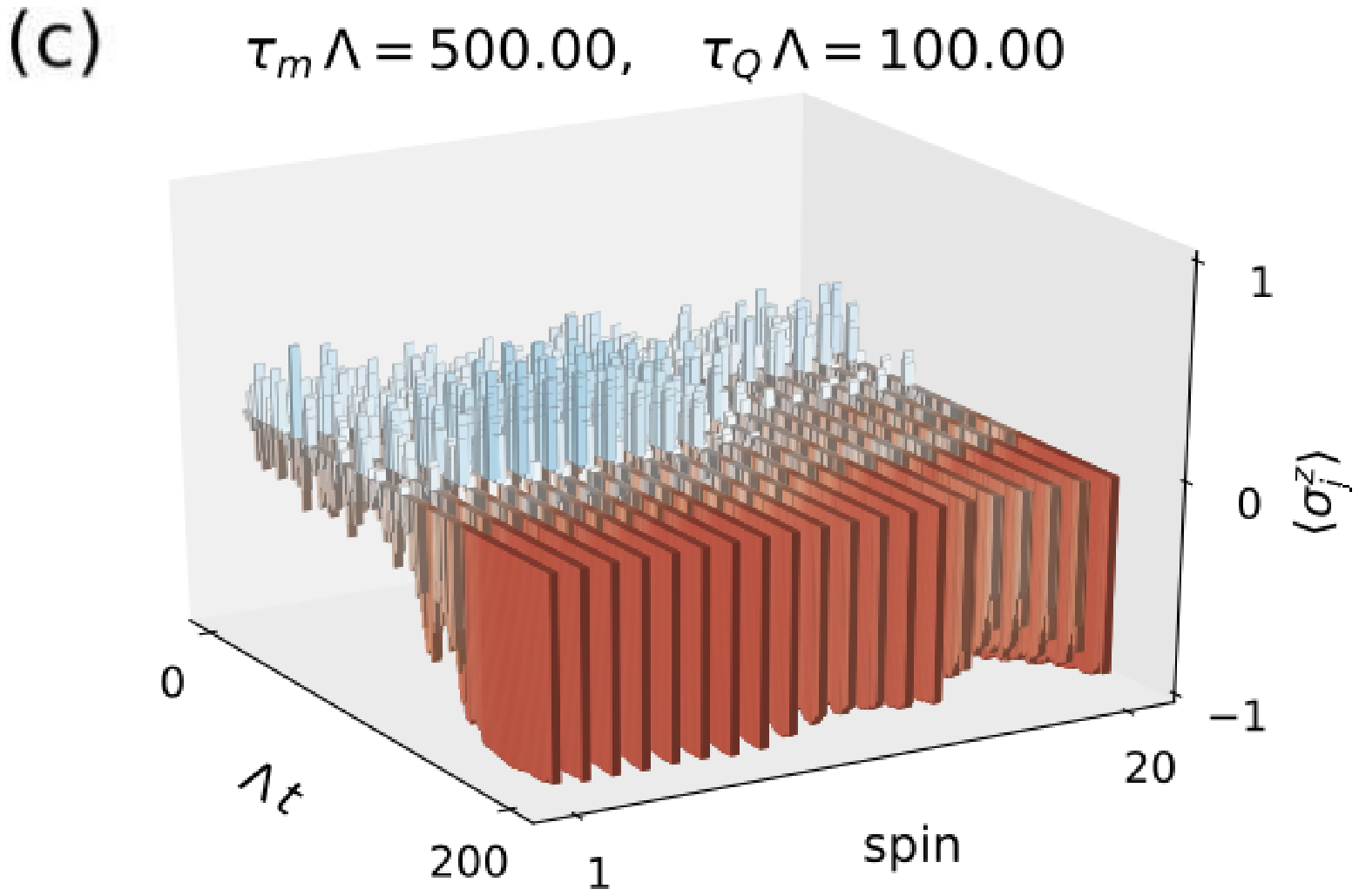}  
\caption{
\label{fig:figure4SM} 
{\bf Monitoring-induced symmetry breaking for a slow quench.}
A system undergoing a slow quench ($1/\tau_Q \ll \Lambda$) tends to remain close to the ground state, but in the absence of monitoring and other sources of symmetry breaking the system cannot select between degenerate ferromagnetic ground states $\ket{ \uparrow \uparrow \dots \uparrow \uparrow}$ and $\ket{ \downarrow \downarrow \dots \downarrow \downarrow}$.
(a) A strong monitoring ($1/\tau_m \sim 1/\tau_Q \ll \Lambda$) of the individual spins generates a symmetry broken state where spins are uncorrelated.
%slightly affects their spin components, barely perturbing the otherwise symmetric evolution. 
(b) As measurement strength decreases, symmetry is broken with a stronger tendency to remain close to the ground state. This favors formation of domains of spins collapsing to stable `up' or `down' configurations.
(c) A weak monitoring ($1/\tau_m \ll \{\Lambda,1/\tau_Q\}$) leads to an almost uniform domain, and a final randomly-selected state that is close to one of the possible symmetry-broken ground states of the ferromagnetic Hamiltonian. 
It is worth noting that the monitoring process affects the state of the system in timescales of order of $\tau_m$~\cite{Bookjacobs2014}, so for this case the effect of the monitoring itself is small in the duration of the simulated experiment.
}
\end{figure*}

\begin{figure*} %[ht]
  \centering
 \vspace{21pt}
        \includegraphics[trim=00 00 00 00,width=0.27\textwidth]{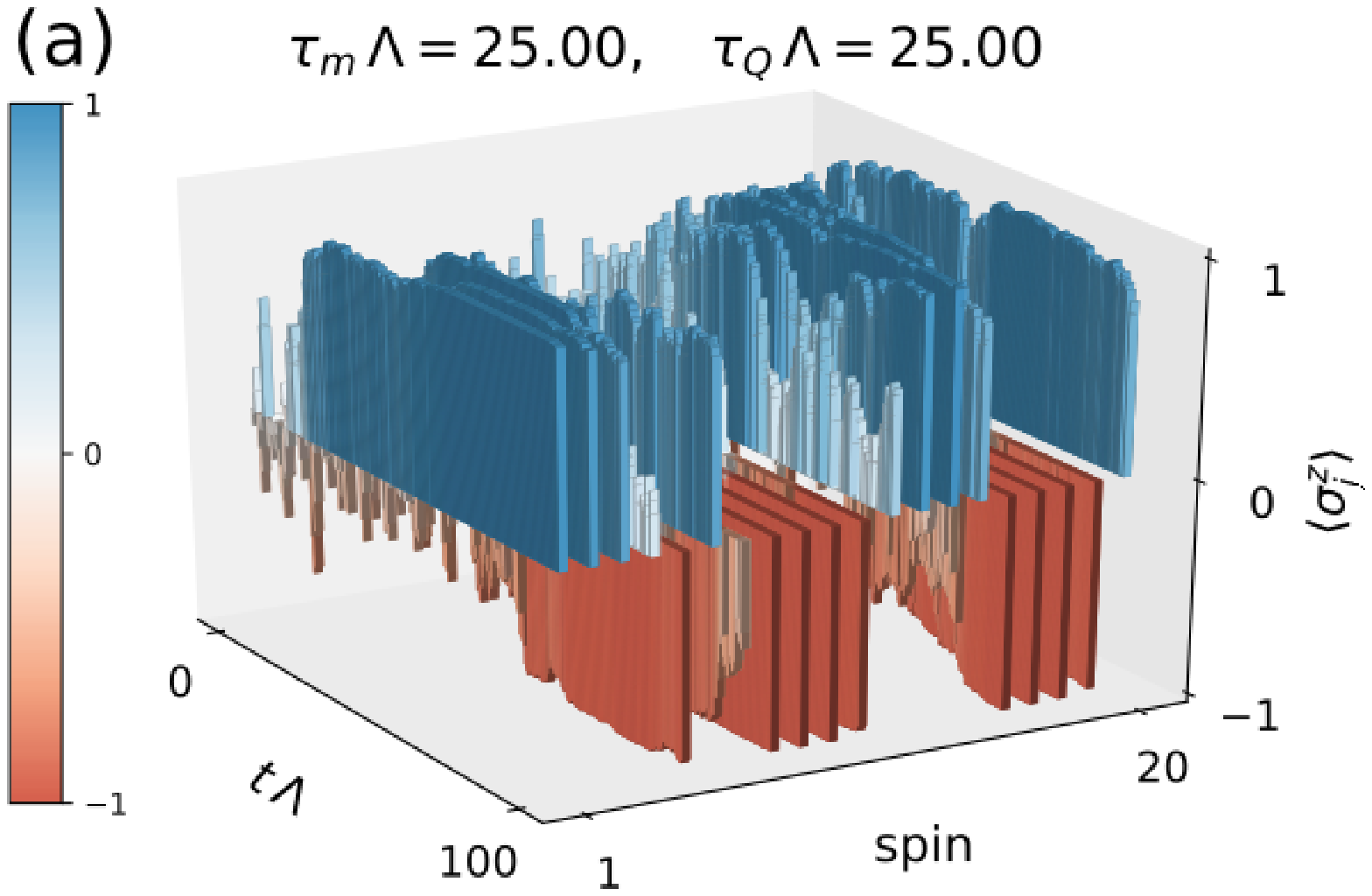}
         \includegraphics[trim=00 00 00 00,width=0.27\textwidth]{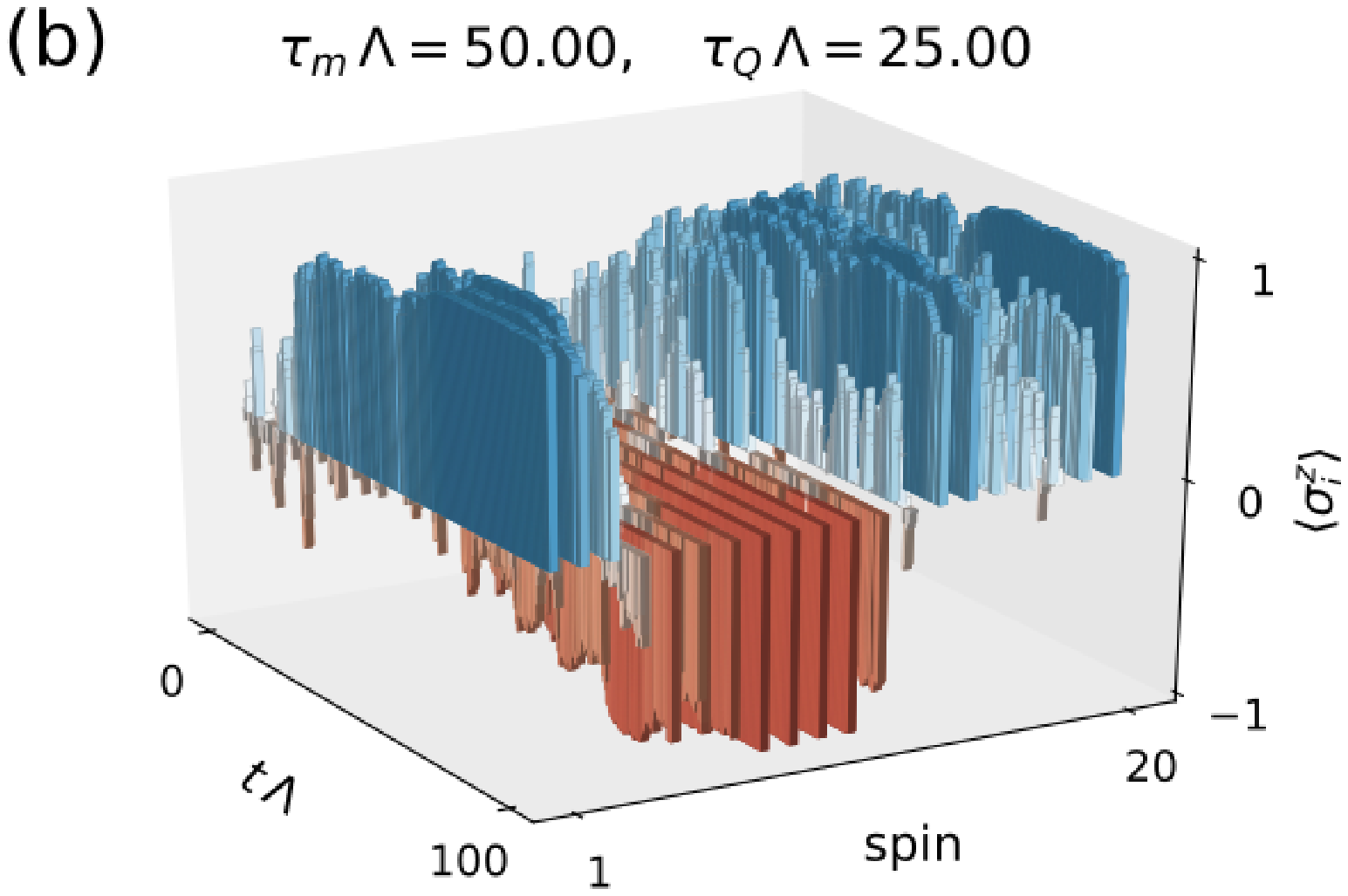}  
       \includegraphics[trim=00 00 00 00,width=0.27\textwidth]{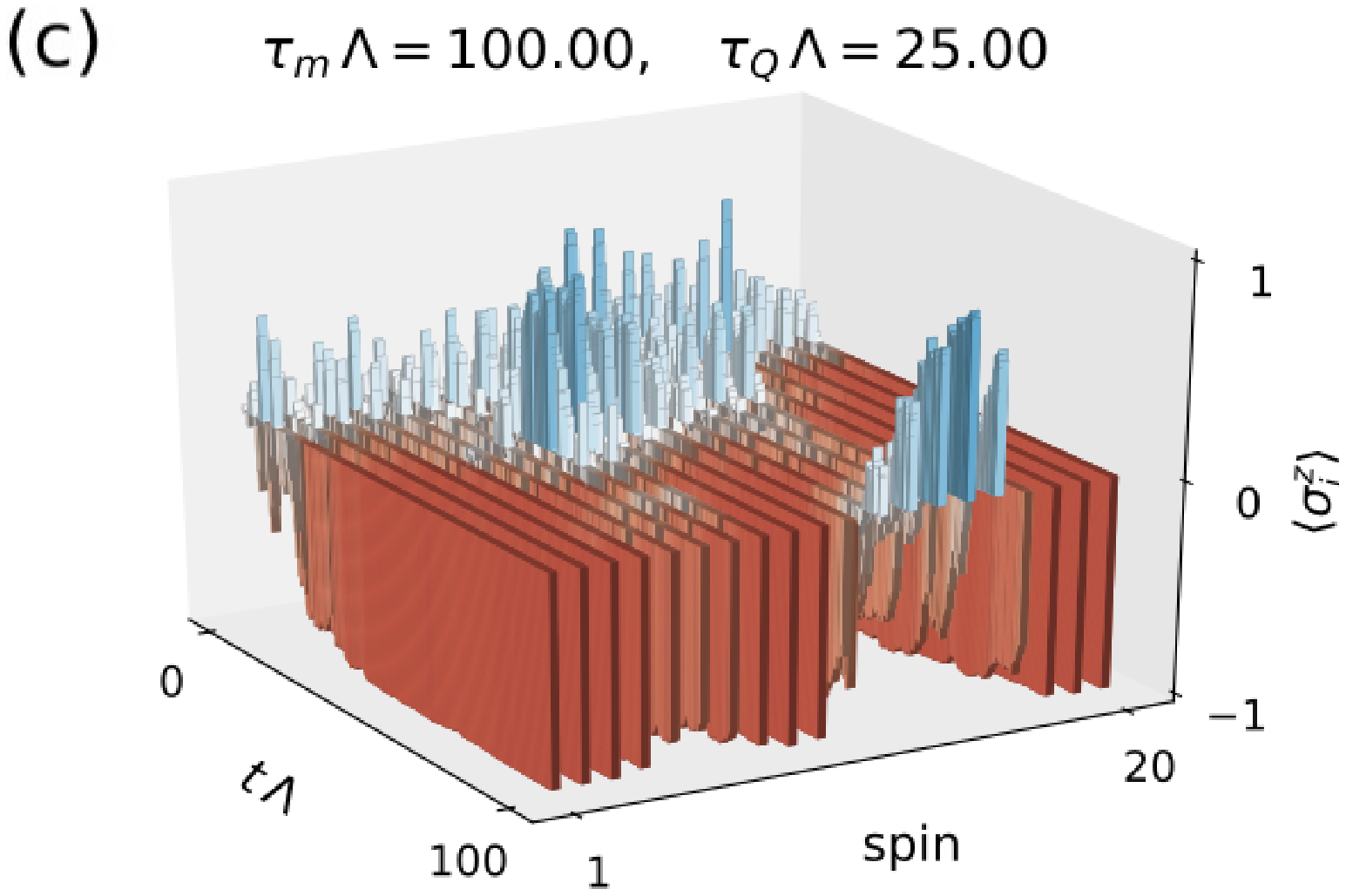}  
\caption{
\label{fig:figure5SM} 
 {\bf Monitoring-induced symmetry breaking for a slow quench.}
For a different regime of the parameters $\{\tau_m,\tau_Q, \Lambda\}$ the same general behavior as in Fig.~\ref{fig:figure4SM} is observed, with formation of larger homogeneous domains as the measurement strength is decreased, in which case the adiabatic dynamics dominates the properties of the resulting symmetry-broken state.
}
\end{figure*}

Without access to measurement outcomes nothing in the dynamics breaks the symmetry between degenerate grounds states of the ferromagnetic Hamiltonian. 
However, the monitoring of the system feeds in a seed of asymmetry. Figures~\ref{fig:figure4SM} and~\ref{fig:figure5SM} illustrate the monitored dynamics for a variety of regimes.
There exists a competition between monitoring back-action, which attempts to single out individual spins, versus the tendency of the system to remain excitation-less due to adiabatic dynamics.  This competition can result in the formation of domains, that is, regions where spins locally align in a given direction, with different definite values of magnetization. The size of such domains depends on the relative strength of the monitoring and of the quench time: while strong monitoring and fast quenches implies uncorrelated individual spins, weak monitoring and slow quenches favors formation of large domains of definite magnetization. In the limit $1/\tau_m \ll \{\Lambda,1/\tau_Q\}$ the system randomly selects one of the otherwise symmetric grounds states $\ket{ \uparrow \uparrow \dots \uparrow \uparrow}$ and $\ket{ \downarrow \downarrow \dots \downarrow \downarrow}$.

%
%\newpage
%
%\clearpage

\subsection{Timescales of purity decay from monitoring-induced symmetry breaking}

SSB in a standard setting, with no continuous measurements, can be induced by  local perturbations, for example to the system Hamiltonian~\cite{MiyaderaPRL2002}.
We consider a seed of SSB to be associated with a stochastic perturbation operator $V$ such that the total system Hamiltonian reads
\begin{align}
H_{\text{st}}(t) = H + \lambda\eta(t)V.
\end{align}
The Hamiltonian $H$ is taken as the ferromagnetic one, and the setting considered is that of a sudden quench from  paramagnetic to ferromagnetic.
Here $\lambda$ is a dimensionless coupling constant characterizing the strength of the perturbation and $\eta(t)$ corresponds to a real Gaussian random process. For simplicity we assume $\langle \eta(t) \rangle=0$ and $\langle \eta(t)\eta(t') \rangle =\delta(t-t')$ which corresponds to the white-noise limit.

The average over many-realizations of the noise makes the dynamics effectively open and described by the Lindblad operator~\cite{ChenuPRL2017,ZhenyuPRL2019} 
\begin{align}
L[\rho_t] &=-i \left[H,\rho_t \right] -  \frac{\lambda^2}{2}[V,[V,\rho_t ]].
\end{align}
 As a result, the evolution of an initial pure state decreases its purity. The time scale in which this SSB seed acts can be identified from the short-time asymptotics of the purity decay and reads
\begin{align}
 \frac{1}{\tau_V}=\frac{\lambda^2}{2}\Delta V^2,
\end{align}
where $\Delta V^2$ is the variance of the SSB seed operator $V$ with respect to the initial state.
This provides the time-scale for SSB induced by the fluctuating operator $V$.
Under continuous quantum measurements the dynamics averaged over many-realizations would be of the form
\begin{align}
\label{eq:Lindblad}
L[\rho_t] &=-i \left[H,\rho_t \right] -  \frac{\lambda^2}{2}[V,[V,\rho_t ]]  -\sum_\alpha \frac{1}{8\tau_m^\alpha} \left[A_\alpha,\left[A_\alpha,\rho_t \right]\right]. 
\end{align}
The short-time purity decay is then set by
\begin{align}
\frac{1}{\tau}= \frac{\lambda^2}{2}\Delta V^2+\sum_\alpha \frac{1}{8\tau_m^\alpha} \Delta A_\alpha^2,
\end{align}
which illustrates the competition between the time scales associated with the stochastic perturbation and the continuous monitoring.

\end{document}